# II. Geometrical framework for thinking about globular proteins: the power of poking

Tatjana Škrbić[1,2], Achille Giacometti[1,3], Trinh X. Hoang[4], Amos Maritan[5], and Jayanth R. Banavar[2]

[1] Ca' Foscari University of Venice, Department of Molecular Sciences and Nanosystems, Venice, Italy

[2] University of Oregon, Department of Physics and Institute for Fundamental Science, Eugene, Oregon, USA

[3] European Centre for Living Technology (ECLT), Ca' Bottacin, Dorsoduro 3911, Calle Crosera, 30123 Venice, Italy

[4] Vietnam Academy of Science and Technology, Institute of Physics, Hanoi, Vietnam

[5] University of Padua, Department of Physics and Astronomy, Padua, Italy

**Correspondence**

Jayanth Banavar
1258 University of Oregon, Eugene, OR 97403-1205, USA
Email: banavar@uoregon.edu

**Abstract**

Recently, we presented a framework for understanding protein structure based on the idea that simple constructs of holding hands or touching of objects can be used to rationalize the common characteristics of globular proteins. We developed a consistent approach for understanding the formation of the two key common building blocks of helices and sheets as well as the compatible assembly of secondary structures into the tertiary structure through the notion of poking pairwise interactions. Here we benchmark our predictions with a detailed analysis of structural data of over 4000 proteins from the Protein Data Bank. We also present the results of detailed computer simulations of a simplified model demonstrating a pre-sculpted free energy landscape, determined by geometry and symmetry, comprising numerous minima corresponding to putative native state structures. We explore the consequences of our model. Our results suggest that symmetry and geometry are a powerful guide to capture the simplicity underlying protein complexity.

**Statement for broader audience**

A poking interaction in a chain causes two parts of a chain to poke towards each other. We use empirical data to establish a two-way link between geometry and chemistry through poking interactions and hydrogen bonds. We show that our simple and tractable geometrical model, superficially unrelated to proteins, can capture the essential features of a rich and incredibly complex protein molecule.

## 1. Introduction

In a recent paper [1], we modeled the backbone of a protein as a chain of $C_\alpha$ atoms with a constant bond length b of approximately 3.81Å, informed by empirical data [2]. Our aim there was to introduce a well-defined and simple model based on geometry that yielded ground states like those of proteins. We presented some preliminary results which showed reasonable accord between geometry and chemistry. We have three goals here. First, we carefully assess the validity of several predictions of the geometrical model by comparing with data from over 4000 native state structures in the PDB. These predictions include backbone related quantities including the tube/coin diameter, the threshold bond-bending angle and numerous constraints pertaining to the relative placement of touching Kepler coins in both a helix and a sheet. Second, to elucidate the nature of the assembly of the tertiary structure, we present the results of detailed computer simulations of a simplified model, without sidechains, but with effective pairwise attraction of a special kind. We present a gallery of nearly degenerate 36 conformations spanning the topologies of commonly observed proteins. And third, we present a brief discussion of some of the consequences of our work. While any model is necessarily wrong, we hope that our model may yet be useful to guide our understanding of proteins, the vital molecules in living cells.

A key concept of our analysis [1] was the notion of poking pairwise interactions (Figure 1) between $C_\alpha$ atoms i and j, located at $\mathbf{r_i}$ and $\mathbf{r_j}$ respectively, satisfying the distance, d, criteria:

$$\begin{aligned} d(i,j) &< d(i,j-1) \\ d(i,j) &< d(i,j+1) \\ d(i,j) &< d(i-1,j) \\ d(i,j) &< d(i+1,j) \end{aligned} \qquad (1)$$

In other words, poking contacts are special in that a given site i is closer to j than it is to its neighbors j+1 and j-1 and likewise for site j – two parts of a chain poke towards each other. We will briefly consider one-way poking contacts later in the paper where not all 4 equations are satisfied but just the top two or the bottom two equations. The building blocks of both helices and sheets arise from the presence of such contacts along with distance and angular constraints linking them and correspond to Kepler-like touching of coins sitting at the $C_\alpha$ locations. These constraints are readily deduced for the simple case of interactions driven by backbone atoms, which are common to all proteins.

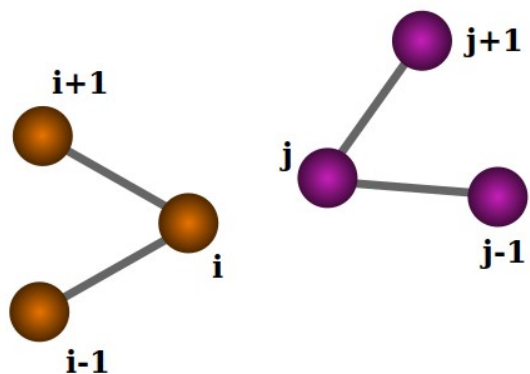

Figure 1: Illustration of two pieces of a chain showing a poking pairwise contact between i and j. i is closer to j than to the two neighbors of j and likewise for j.

For the assembly process of the secondary motifs into the tertiary structure, heterogeneous sidechains are involved in the contacts, and touching entails non-uniform objects with imperfectly known geometries coming together and nestling with each other. We suggested [1] that the complex many-body interaction of multiple sidechain atoms nestling together and expelling the water from the hydrophobic core can be approximately captured by means of additive poking pairwise interactions within a specific range and which are *not* involved in building block formation. A poking contact plays an essential role in both helix and sheet formation, and it is the simplest emergent pair-wise interaction that one may postulate when there is little specific knowledge [1].

We considered the consequences of a constructive hypothesis that the common building blocks of protein native state conformations satisfy certain Kepler symmetry constraints. Because of chain tethering, one ought to consider the objects tethered along the chain not as isotropic spheres but, in the simplest case, as two coins. This is because of a special tangent direction at each site. For a helix, the normal directions of the two coins at a given site indicate the directions from the site to its nearest neighbors (Figure 2b). A strand is associated with two parallel straight axes, one going through (i-2,i,i+2) and the other going through (i-1,$M_i$,i+1), where $M_i$ is the mid-point of (i-1) and (i+1). We suggested that the two axes associated with the i-th $C_\alpha$ atom are represented by one coin at site i and the other rigidly translated to $M_i$ (Figures 3a and 3b). All backbone coins are taken to have a constant radius $\Delta$, equal to the radius of curvature of the Kepler helix [1].

We sought building block conformations [1] in which pairs of coins just touched each other edge on edge or held hands in a systematic manner with the right geometrical constraints. This requirement was met perfectly in two systematic chain geometries (Figures 2b, 3a, and 3b) – a Kepler helix with the two coins at i just touching one coin each at (i±3) and two possible arrangements of zig-zag strands in phase and tracking each other in a Kepler sheet. The two Kepler sheets correspond to parallel and antiparallel strands with separation between the pairing axes of neighboring strands being 2$\Delta$.

Our theory [1] predicts that for a bond length of around 3.81Å, the Kepler helix has a coin radius $\Delta$ of around 2.63Å. Furthermore, steric constraints dictate that the bond-bending angle of the chain is *nominally* larger than $\theta_{min}$ around 87.3°. In the continuum limit, our model becomes a tube of non-zero radius $\Delta$. The non-zero $\Delta$ provides the space to fit in the atoms of the backbone atoms within the tube.

Figure 2a depicts the arrangement of coins, each having a radius Δ, in a continuous helix [3]. The continuum approximation does not pick out a characteristic length scale because the bond length approaches zero in this limit. Here we have, for consistency, chosen the continuum tube to have a radius Δ=2.63Å to match that of the discrete Kepler helix (Figure 2b). In the continuum case, the central coin in the figure just touches two other coins. In contrast, in the discrete case, there are two central coins, each just touching one other coin. The continuous and discrete helices turn out to have virtually the same, *though not identical*, geometries measured by the pitch to radius ratio P/(2πR) ~ 0.4. Here P is the pitch of the helix and R its radius.

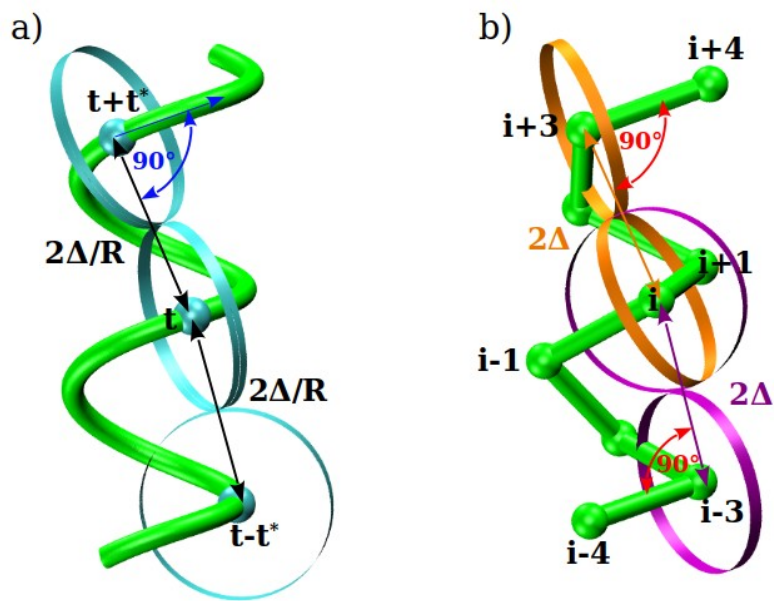

Figure 2: a) Continuous helix defined by two dimensionless ratios η=P/(2πR) and Δ/R = 1+η$^2$, P and R are the helix pitch and radius. The tube (coin) radius is denoted as Δ. The helical curve is shown in green. The three cyan points along the helical curve are separated by successive turn angles of t$^*$ ~ 302.5° and, unlike in the discrete helix, each is associated with just one cyan coin having a dimensionless radius Δ/R. This is because the directions from point i to both its neigh-

bors along the helix coincide in the continuum limit. The coin in the middle, just touches the other two coins with the distance between coin centers being exactly equal to 2Δ/R. Also, the straight line that connects successive cyan points is perpendicular to the tangent of the helical curve at those points. These geometrical conditions satisfy the Kepler touching condition. b) Kepler helix having bond length b. For a discrete chain, there are now two distinct coins at site i (in orange and in purple). This is because the (i-1,i) and the (i,i+1) directions do not coincide any longer. The figure shows one of two coins at sites i-3 (in purple) and i+3 (in orange). These two just touch the two coins at bead i. The pair of orange coins touch each other as do the pair of purple coins. The distances of 2Δ (the coin diameter) and the angles of 90° characterizing the geometrical conditions of touching are indicated. Every pair of non-contiguous coins that do not touch is farther than 2Δ from each other and therefore cannot intersect. (i,i+3) is a poking contact because i is closer to i+3 than it is to i+2 and i+4, and i+3 is closer to i than it is to i-1 and i+1. The same logic can also be applied to the formation of sheets (Figures 3a and 3b).

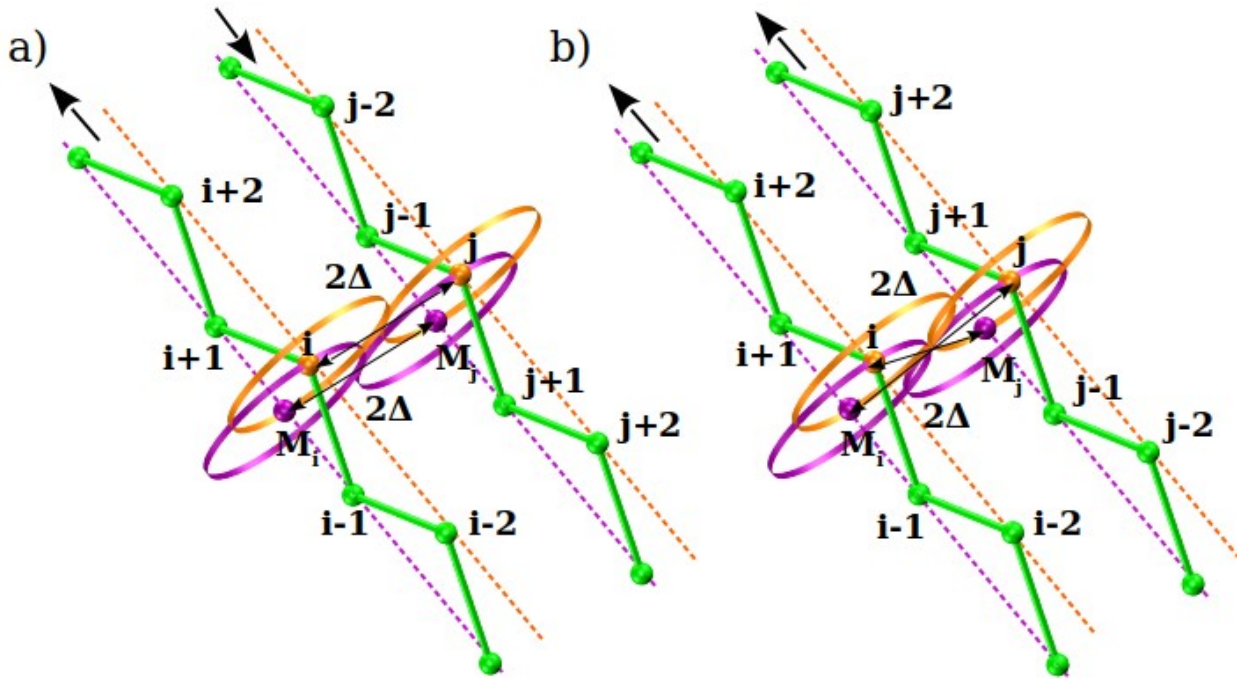

Figure 3: Two possibilities for the coordinated handholding for a pair of identical ideal strands. Each strand has two axes. a) shows a pair of antiparallel strands a distance $2\Delta$ apart with associated pairs of coins (shown in orange and purple) of radius $\Delta$ at sites i and j that are oriented along the orange and purple axes respectively. The purple axis of the left strand goes through sites (i-1, $M_i$, i+1) whereas the orange axis passes through the points (i-2, i, i+2). The orange coins of i and j touch each other as do the purple coins centered at $M_i$ and $M_j$. The (i-j) distance is $2\Delta$. b) shows a distinct conformation of parallel strands. The orange coin at i now just touches the purple coin at $M_j$ and the orange coin at j just touches the purple coin at $M_i$. The (i-$M_j$) distance is now $2\Delta$ and the (i,j) distance is smaller.

One must recognize that Kepler chain conformations may have nothing to do with protein structures. Kepler's handholding and poking and nestling are geometrical schemes, which can be studied simply. Because of the venerable history in the protein field, dating back to Pauling [4,5], Ramachandran [6], Richards [7-9], and Rose [10-13] of packing and space-filling, one may hope that Kepler structures resemble protein structures. This was also seen in earlier work on the optimal shapes of compact strings more than two decades ago [3]. Unlike in a protein, where there are myriad interactions, including steric constraints, covalent bonding, hydrogen bonding, dipolar interactions, van der Waals forces, electrostatics, and hydrophobicity [14-17], the Kepler model is ultra-simple. Here we carry out detailed comparisons between the predictions of theory and empirical data from the building blocks of over 4000 native state structures and find good accord. To complete our analysis, we have also carried out extensive simulations of a simplified model [1] (see Materials and Methods). Our primary goal here is to explore whether the results of our geometrical framework are in quantitative accord with protein data. We will discuss the consequences of our findings in the concluding section of this paper.

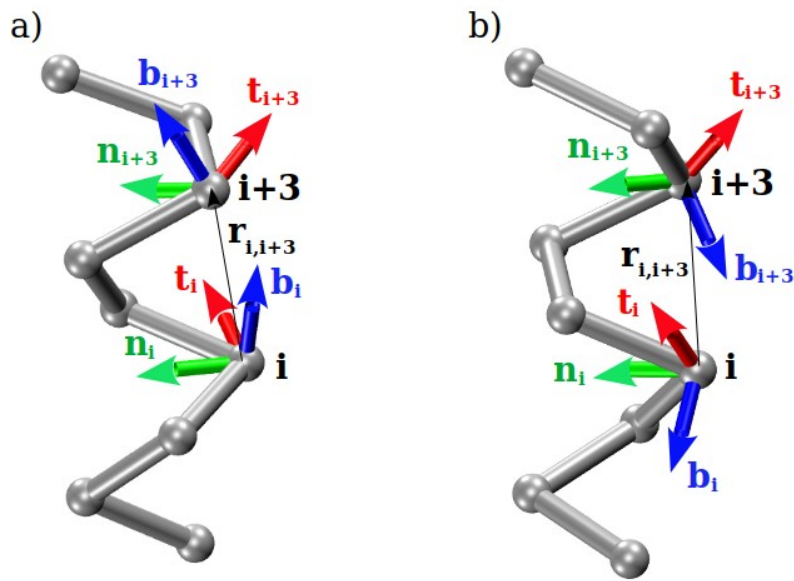

Figure 4: a) Right-handed and b) left-handed Kepler helix. Protein helices are predominantly right-handed because the constituent amino acids themselves are left-handed. Chiral symmetry breaking originates from steric clashes of oxygen backbone atoms with the side chain atoms in an extended left-handed helix [6,10]. In our simulations, we break this symmetry by hand. a) Right-handed helix and the local Frenet frames [19] ($t_i,n_i,b_i$) and ($t_{i+3},n_{i+3},b_{i+3}$) at the locations of beads i and i+3. The normal direction at i is determined by drawing a circle through (i-1,i,i+1) and joining i to the center of the circle. The tangent direction is along the line joining i-1 and i+1. The (tangent,normal,binormal) unit vectors form a right-handed Cartesian coordinate system. Note that the sign of the scalar products $t_i \cdot b_{i+3}$ and are positive in the right-handed Kepler helix and equal to ~0.241. [This corresponds to an angle between $t_i$ and $b_{i+3}$ vectors (as well as an angle between $b_i$ and $t_{i+3}$ vectors) of ~76°]. b) Left-handed helix and the local Frenet frames ($t_i,n_i,b_i$) and ($t_{i+3},n_{i+3},b_{i+3}$) at the locations of beads i and i+3. The scalar products $t_i \cdot b_{i+3}$ and $b_i \cdot t_{i+3}$ are negative in the left-handed helix and equal to ~-0.241. These dot products are a convenient way of diagnosing chirality and enforcing chiral constraints on helices. In the Kepler right handed helix, the angle between the (i-1,i) direction and the $r_{i,i+3}$ vector is 90°. From simple geometry, the corresponding angle between the $t_i$ and $r_{i,i+3}$ unit vectors is ~77.7° and their scalar product is ~0.213. We use this condition as a constraint in our simulations to obtain the Kepler helix in lieu of using the condition that $r_{i,i+3}$ is perpendicular to the (i-1,i) direction.

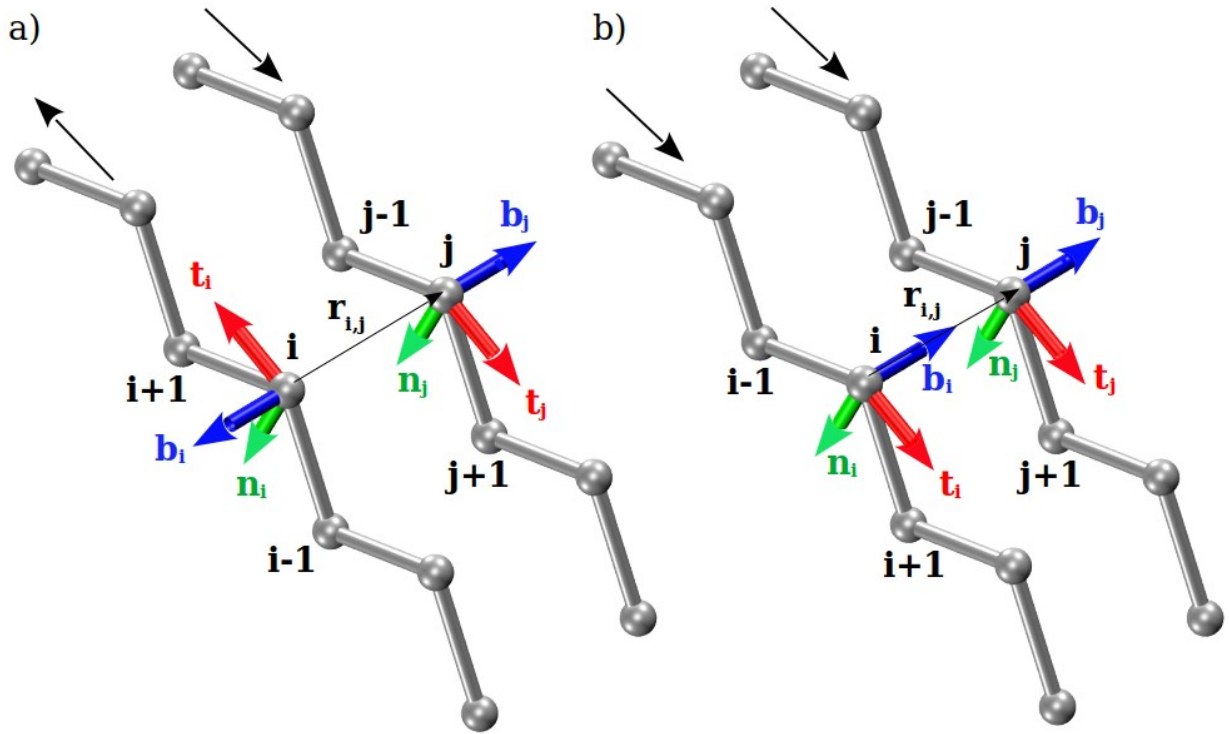

Figure 5: a) Antiparallel and b) parallel arrangements of Kepler strands. Local Frenet frames [19] ($\mathbf{t}_i,\mathbf{n}_i,\mathbf{b}_i$) and ($\mathbf{t}_j,\mathbf{n}_j,\mathbf{b}_j$) at the locations of beads i and j. The normal direction at i is determined by drawing a circle through (i-1,i,i+1) and joining i to the center of the circle. The tangent direction is along the line joining i-1 and i+1. The (tangent,normal,binormal) unit vectors form a right-handed Cartesian coordinate system. Note that in idealized strands scalar products $\mathbf{n}_i \cdot \mathbf{r}_{ij}$ and $\mathbf{n}_j \cdot \mathbf{r}_{ij}$ are 0.

## 2. Materials and Methods

### 2.1 PDB data and computer simulations

We follow the Methods described in Ref. [1]. Our data consists of 4,391 PDB structures, a subset of Richardsons' Top 8000 set [21] of high-resolution, quality-filtered protein chains (resolution < 2Å, 70% PDB homology level), that we further filtered to exclude structures with missing backbone atoms, as well as amyloid-like structures. The presence of hydrogen bonds was identified using DSSP [22].

We carried out a detailed study within our protein data set of (i,i+3) poking contacts, which we divided into two classes. The first class is that of an embedded contact, the kind that one would expect to find in the interior of a well-formed helix. For these we require that (i-1,i+2), (i,i+3), (i+1,i+4) form a triplet of poking contacts. The second class of isolated poking contacts comprises (i,i+3) poking contacts, while the two neighbors do not form even one-way poking contacts. We remind the reader that in Eq. (1), for one-way poking contacts, just the top two or the bottom two equations are satisfied. We have identified 176,711 embedded and 25,893 isolated poking contacts of the type (i,i+3) in our data set consisting of 4,391 globular proteins.

To characterize handholding in Kepler sheets, we have analyzed 13,442 parallel poking contacts and 28,118 antiparallel ones of the type (i,j) with j>i+3 in our data set. We eliminate edge effects by choosing contacts in the interior of the sheet by requiring that an (i,j) poking contact in a parallel arrangement is accompanied by poking contacts between (i-2,j-2) and (i+2,j+2). For the antiparallel arrangement, the (i,j) poking contact is required to have partner pairs of beads (i-2,j+2) and (i+2,j-2) that have poking contacts as well.

In our earlier paper [1], we presented data showing that hydrogen bonds, which provide scaffolding for the protein building blocks, were associated with poking contacts. Here we will show that the reverse holds and that pok-ing contacts associated with the backbone atoms are generically associated with hydrogen bonds. This result demonstrates the two-way correspondence between geometry and chemistry.

As detailed in [1], we have employed two complementary Monte-Carlo techniques in our simulations to obtain ground state conformations of our emergent space-filling model: microcanonical Wang-Landau (WL) simulations [23] and replica exchange (RE) (or parallel tempering) canonical simulations [24]. We used both methods to check for consistency and find conformations with low energies. The acceptance probability in the WL approach provides a convenient guide to explore less populated (and thus low-lying) energy states and flattens the energy histograms over the course of the simulations. The RE approach consists of canonical simulations that are conducted in parallel over a wide range of temperatures that bracket the 'transition temperature' between the folded and unfolded states, while concentrating in the low temperature region to search for low-lying states. Each simulation provides a replica of the system in thermal equilibrium. The swapping of replicas allows for rapid search. Both methods employed standard local moves including crankshaft, reptation, and endpoint moves along with the non-local pivot move.

Our simulation model has the following features [1]:

- We impose a constraint on the local bond-bending angle that it cannot be smaller than $\theta_{min} \sim 91°$. We also require that no pair of $C_\alpha$ atoms can approach each other within a hard-core distance (of order of the van der Waals radius of a glycine residue of 4.5Å [18]) from each other. The results of our analysis are robust upon varying both these quantities.

- Poking contacts within a helical α-basin are assigned a happiness energy, $-E_\alpha$. These intra-helical pairwise poking contacts of the (i,i+3) type are required to satisfy the correct constraints. We implement these requirements by imposing soft constraints on the values of the dot products $\mathbf{t_i} \cdot \mathbf{b_{i+3}}$ and $b_i \cdot t_{i+3}$ as well as the dot products $t_i \cdot r_{i,i+3}$ and $t_{i+3} \cdot r_{i,i+3}$. Here $t_i$ and $b_i$ represent the tangent and binormal vectors at bead i in the local Frenet system of coordinates [19] and $r_{i,i+3}$ is the vector connecting beads i and i+3. Specifically, we assign no α-basin reward unless the dot products $t_i \cdot b_{i+3}$ and $b_i \cdot t_{i+3}$ as well as $t_i \cdot r_{i,i+3}$ and $t_{i+3} \cdot r_{i,i+3}$ lie in appropriate ranges that are deduced from the *righthanded* Kepler helix. The dot products $t_i \cdot b_{i+3}$ and $b_i \cdot t_{i+3}$ both need to lie between +0.156 and +0.325 allowing for ±5° tolerance around the ideal angle of ~76° between the corresponding vectors. We note that here we overtly break chiral symmetry by imposing positive values on the dot products $t_i \cdot b_{i+3}$ and $t_{i+3} \cdot b_i$ to ensure that the preferred helix is righthanded (see Figure 4). Likewise, the dot products $t_i \cdot r_{i,i+3}$ and $t_{i+3} \cdot r_{i,i+3}$ need to lie in the range between +0.127 and +0.297, permitting a ±5° tolerance around the ideal angle of ~77.7° between the corresponding vectors in the Kepler helix. We note that the dot products $t_i \cdot r_{ij}$ and $t_j \cdot r_{ij}$ do not depend on the chirality of the Kepler helix (see Figure 4).

- Poking contacts in the β-basin are assigned a happiness energy, $-E_β$. The zigzagging of an individual strand is ensured [20] by considering the relative orientations of $n_i$ and $n_{i+1}$, where $n_i$ is the normal vector at site i (Figure 5). In addition, we require a nearly perpendicular orientation of the connecting vector between i and j, $r_{ij}$, with both $n_i$ and $n_j$, to account for the parallelism of the paired axes in a β-sheet. We do this by allowing for ±5° tolerance around the ideal angle of 90° between vectors $n_i$ and $r_{ij}$, as well as $n_j$ and $r_{ij}$ vectors (Figure 5), while the zigzagging is imposed by requiring that the dot products $n_i \cdot n_{i-1}$ and $n_i \cdot n_{i+1}$ of an individual strand be smaller than -0.5 ensuring that the angle between successive normals is at least 120°. We remind the reader that for ideal strands this angle is 180°, with the scalar product being -1. The geometrical rules we employ in the β-basin do not distinguish between parallel and antiparallel arrangement of strands. Thus, as noted in our earlier paper [1], our model is deliberately simplified and does not consider the sub-lattice symmetry of each strand. The net result is that structures obtained in our simulations are a bit more idealized than they ought to be and there is no squeezing or swinging. Furthermore, the distinct geometries of parallel and antiparallel strands are not captured in our simplified model.

- Finally, we reward poking contacts that do not fall in the α- or β-basin with a happiness energy, -$E_\gamma$, if they are within a range 12Å. We have verified that our results are independent of the precise value of this attraction range within ±25%. Contacts in the γ-basin are responsible for facilitating the assembly of α-helices and β-sheets into the tertiary structure of the protein. We will analyze the sensitivity of our results later in the paper on the precise value of the crucial $E_\gamma$ parameter.

## 3 Results

### 3.1 Protein structure

We will state the predictions of our theoretical framework [1] and compare them quantitatively to empirical data. We begin with an observation and two general predictions.

*Poking contacts:* A key ingredient of our analysis is the concept of poking contacts defined in Equation (1). Figure 6 presents a histogram of the (i,j) physical distance in poking contacts (i,j) of the proteins in our data set. This is a measure of the distance between $C_\alpha$ atoms, i and j, making a poking contact. There are two pronounced minima in the distribution corresponding to separations of 6Å and 12Å, which correspond roughly to the backbone coin diameter and around twice the maximum side chain protrusion length respectively. This graph with a two-shell structure underscores the two distinct mechanisms within a protein facilitating Kepler touching and nestling, the first involving backbone-backbone interactions and the second capturing the nestling of heterogeneous sidechains through effective interactions between the backbone atoms.

*Coin/tube radius:* Our theory [1] predicts a coin/tube radius $\Delta$ of around 2.63Å. This is the length scale which is equal to the radius of curvature of the continuous Kepler helix and is given by $R_{curv} = R(1+\eta^2)$, where $\eta=P/(2\pi R)$, P is the helix pitch, and R is the helix radius. It is also predicted to be equal to half the distance between sites separated by three along a Kepler helix. The unique geometry of the Kepler helix and the coin radius is determined by solving these equations along with ensuring that i and i+3 form a poking contact in a helix. Figure 7 shows a representation of the hemoglobin protein showing that almost all backbone atoms are nicely contained within the tube of the predicted radius.

*Local bond bending angle:* Theory predicts [1] that, on average, the minimum bending angle, $\theta_{min} = 2 \sin^{-1}(\Delta/(2b)) \sim 87.3°$. There can of course be exceptions because of the twenty varieties of amino acids including small ones like glycine, which are often associated with backward bends. Figure 8 depicts a histogram of bond bending angles $\theta$ in our protein data set. There are two observations. The first is that only a small percentage (~2.8%) of local bending angles is smaller than $\theta_{min}$. Also, the predicted bond bending angle of the Kepler helix (indicated by the blue arrow) occurs close to the most prominent peak in the distribution.

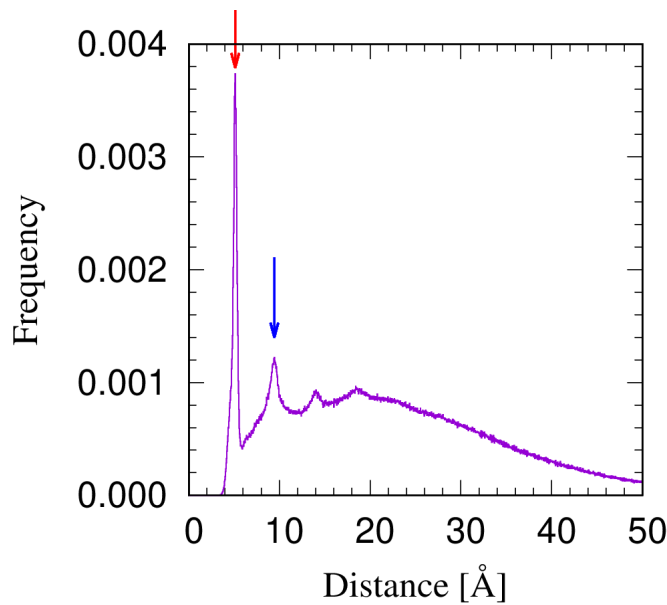

Figure 6: Histogram of mutual distances between 7,278,619 poking contacts in our data set comprising 4,391 globular proteins (purple line) showing a highly pronounced shell structure. The vertical red arrow indicates the location of the peak in the first shell of poking contacts at 5.12Å that finishes sharply at 6.0Å. The contacts in the first shell include touching Kepler coins theoretically predicted to occur at 2Δ = 5.26Å. The vertical blue arrow indicates the location of the peak in the second shell of poking contacts at 9.44Å. The second shell finishes around 12Å, which roughly corresponds to the twice the size of the largest amino acid. We use the latter number for the range of sidechain mediated poking contacts responsible for nestling and assembly of secondary structures.

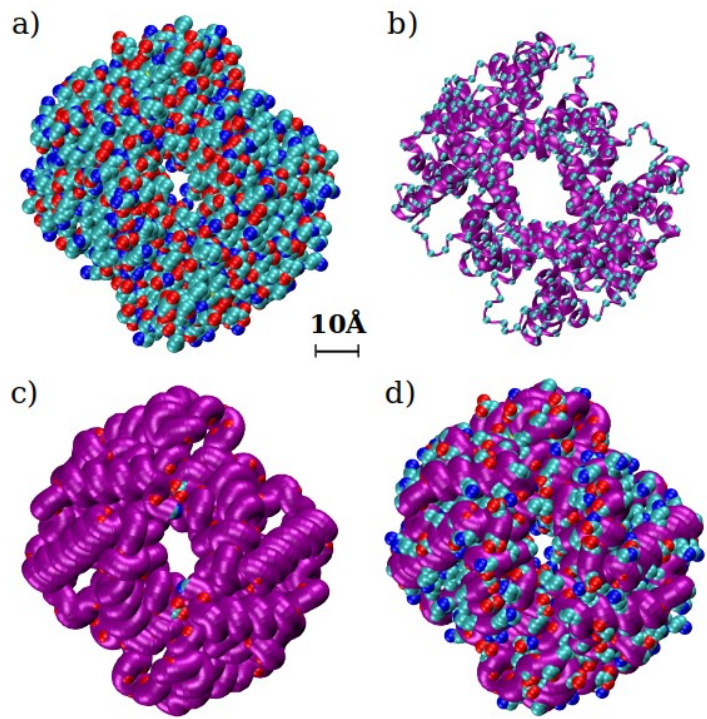

Figure 7: a) Native state of hemoglobin (PDB code: 1A3N) in the CPK representation [4,5] in which all heavy atoms of the protein backbone and its side chains are represented as spheres with radii proportional to their respective van der Waals atomic radii. Color code: carbon (cyan), oxygen (red), nitrogen (blue), and sulfur (yellow). b) Simplicity underlying complexity. The hemoglobin native state structure is shown in ribbon representation (in purple) with cyan spheres, shrunk in size for the sake of clarity, at the positions of the $C_\alpha$ atoms. c) The hemoglobin structure shown in a tube representation (also in purple) with the tube diameter chosen to be the theoretically predicted value of 5.26Å. The relatively few backbone oxygen atoms (red spheres) not entirely enclosed by the tube are visible. d) The same tube representation but this time depicting the backbone and side chain atoms not fully enclosed by the tube.

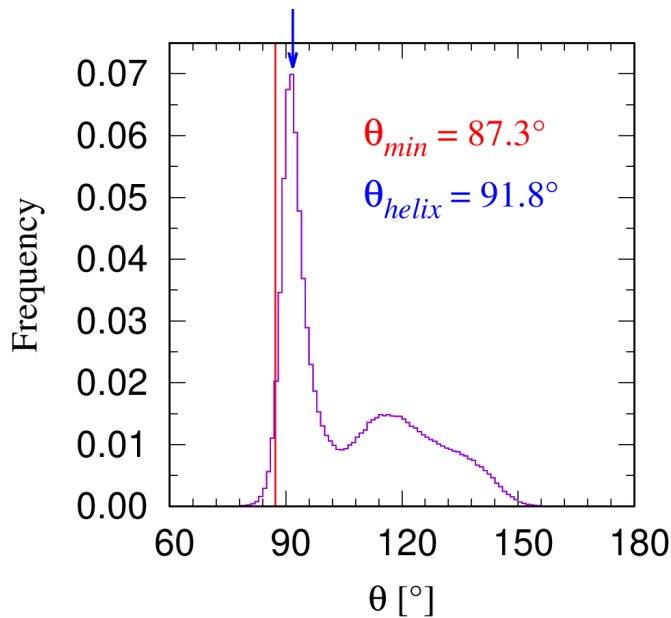

Figure 8: Histogram of 970,896 bond bending angles θ in our data set of 4391 globular proteins (purple line). The vertical red line indicates the theoretically predicted $\theta_{min}$ ~ 87.3°. Approximately 2.8% of bond bending angles have values smaller than $\theta_{min}$. Around 41% of these small bond bending angles are in helices, while the rest are in loops. The vertical blue arrow indicates the predicted θ value for the Kepler helix, which is close to the most pronounced peak in the distribution.

## 3.2 Kepler helix

Our model is coarse-grained, and we represent each residue by just its backbone $C_\alpha$ atom. In contrast, Ramachandran [6] and Rose [10-13] have done vital work looking at steric constraints imposed by all the backbone atoms participating in the planar peptide bond.

Ramachandran showed [6] that two of the important sterically allowed regions in (φ,ψ) conformation space correspond to the helix and the sheet. Thus, steric constraints, without even a need to consider hydrogen bonds, is consistent with the two building block conformations of a helix and a sheet. We also find this result in our Kepler analysis. However, what eludes our coarse-grained approach is the clear preference for the right-handed helix over the left-handed helix [6,10]. This chiral symmetry breaking arises from backbone atom clashes in sustained left-handed helices when one considers the left-handedness of the constituent amino acids. In contrast, our model exhibits chiral symmetry, and it must be broken by hand, as we will, to get protein-like behavior.

Rose and colleagues [25] demonstrated two key results: the breakdown of the Flory isolated pair hypothesis [26] underscoring the influence of steric constraints to beyond nearest neighbor residues and, even more fundamentally, the digital nature of proteins highlighted by a steric clash when a string of α α α α… residues is immediately followed by a β residue.

We proceed now to a study of poking (i,i+3) contacts in our data base obtained from the PDB. These contacts are critical for handholding in a helix, and they are purely geometrical in nature. We will distinguish between two classes of (i,i+3) poking contacts (see Methods for a description of embedded and isolated (i,i+3) contacts) and highlight the differences in their statistical properties in Figure 9. Most such poking contacts are associated with a hydrogen bond between (i,i+3), or (i-1,i+3), or (i,i+4) residues. While this is not a surprise for poking contacts in the helix interior, it is not obvious that this should be the case for an isolated poking (i,i+3) contact. We identify 25,893 totally isolated poking contacts in our database, of which ~80% are effectively hydrogen bonded between (i,i+3) $C_\alpha$ atoms, ~12% between (i-1,i+3) $C_\alpha$ atoms, and ~2% between (i,i+4) $C_\alpha$ atoms. Note that our description of a protein chain is based on a coarse-grained $C_\alpha$ representation of the backbone and does not consider C=O or N-H groups between which hydrogen bonds form.

A mere ~6% are completely devoid of hydrogen bonds as determined by the DSSP [22]. Unlike the isolated contacts, less than ~ 0.05% of the 176,711 embedded poking contacts are not associated with hydrogen bonding. This provides a remarkable link between chemistry [27] and geometry.

More than ~ 96% of (i,i+3) fully embedded poking contacts correspond to a DSSP [22] categorization of (H,H), while around ~ 3.5% of such pairs are classified as (H,X) pairs by DSSP (X=T,I,S,C,G), but only around ~0.2% of such pairs are classified by (X,H) pairs by DSSP (X=T,I,S,C,H) . This underscores the asymmetry between the beginning and ending of a helix [28,29].

There are marked differences in the histograms of the (i,i+3) distances for embedded and isolated poking contacts (Figure 9a). The result for embedded contacts is compatible with the theoretical prediction of 2Δ = 5.26Å for the Kepler helix [1] within error bars. The isolated poking contacts are more spread out but peak around a value close to 2Δ. Figure 9b shows the histograms of the angles (i-1,i,i+3) and (i,i+3,i+4), which are predicted to be 90° for the Kepler helix. There is good accord within error bars for the embedded poking contacts. The isolated poking contacts, on the other hand, exhibit angles, which are less constrained and much more spread out.

Figure 9c shows histograms of the scalar products $\mathbf{t}_i \cdot \mathbf{b}_{i+3}$ and $\mathbf{t}_{i+3} \cdot \mathbf{b}_i$, where $\mathbf{t}$ and $\mathbf{b}$ represent the tangent and binormal vectors at a given site (Figure 4a). These quantities provide an excellent diagnostic of helix chirality. The empirical data for the embedded poking contacts is consistent with the theoretical prediction of +0.241 (corresponding to an angle ~ 76°) for the dot products in the Kepler helix and indicated by the red vertical line. The positive value of the dot products corresponds to a right-handed helix (a left-handed helix would have negative dot products of the same magnitude) and can be used to determine or enforce helix chirality. In contrast, an isolated poking contact can be left-handed but cannot consolidate into a sustained left-handed helix as mentioned earlier.

Figure 9d shows histograms of the scalar products $\mathbf{t}_i \cdot \mathbf{r}_{i,i+3}$ and $\mathbf{t}_{i+3} \cdot \mathbf{r}_{i,i+3}$. Here $\mathbf{r}_{i,i+3}$ is the vector joining i with i+3. Whereas the handholding requirement is that the line joining (i,i+3) is perpendicular to both (i-1,i) and (i+3,i+4), these two latter directions do not coincide with the tangent vectors at i and i+3. Straightforward geometry shows that, for the Kepler helix, the scalar products $\mathbf{t}_i \cdot \mathbf{r}_{i,i+3}$ and $\mathbf{t}_{i+3} \cdot \mathbf{r}_{i,i+3}$ ought to be +0.213 (corresponding to an angle ~77.7°). We use these conditions in our simulations to reward and promote a Kepler helix. Again, the constraint for an isolated poking contact is much more relaxed.

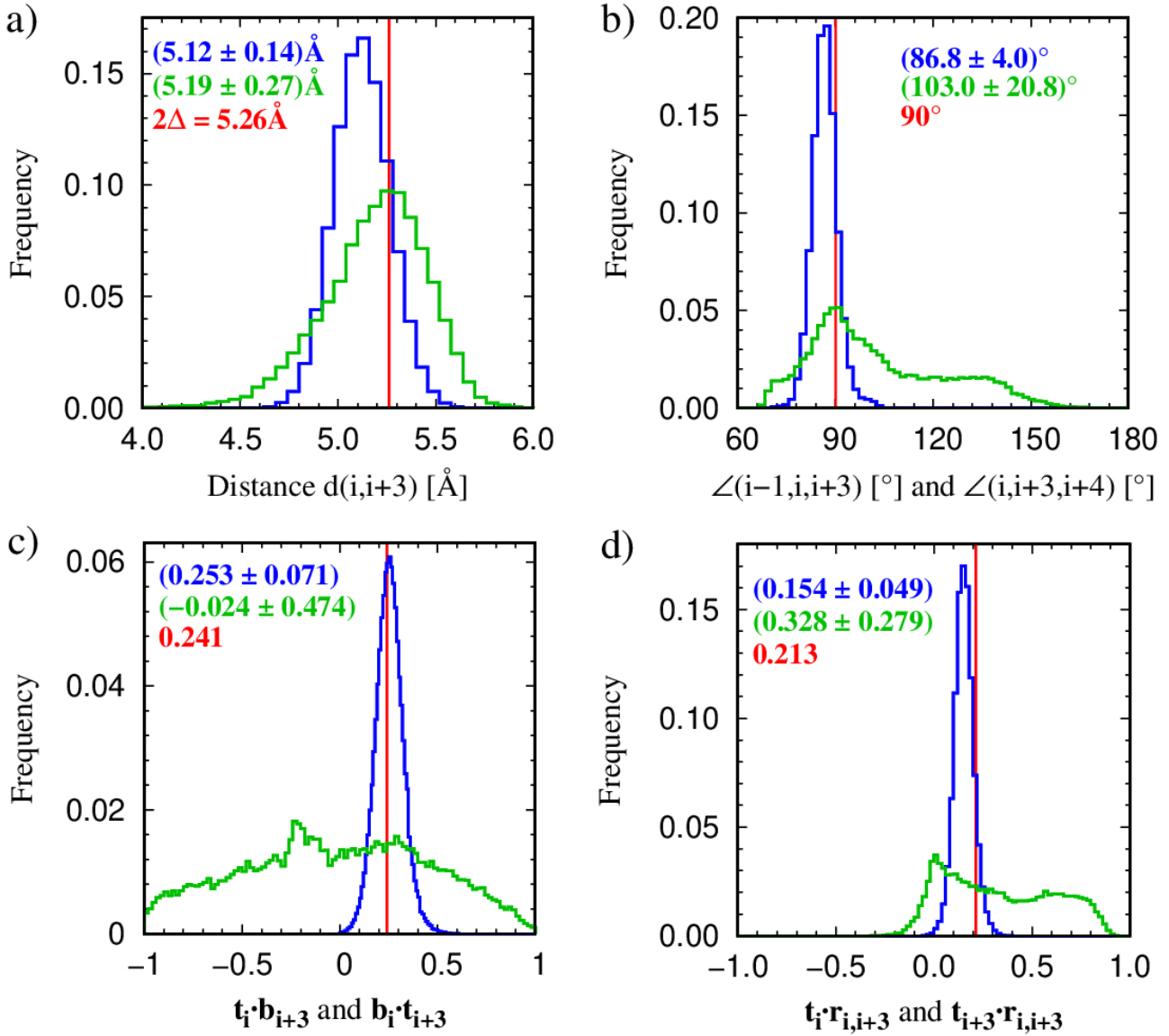

Figure 9: Geometrical characteristics of embedded and isolated local poking contacts along the chain of the type (i,i+3). An embedded (i,i+3) poking contact is defined as one for which both the neighboring pairs (i-1,i+2) and (i+1,i+4) are also poking contacts. In this situation, the (i,i+3) contact is most likely embedded within a helix. A (i,i+3) poking contact is denoted as isolated when neither neighboring pair, (i-1,i+2) or (i+1,i+4), is poking even in an asymmetric manner. An isolated poking contact is thus truly isolated. We have identified 176,711

embedded and 25,893 isolated poking contacts of the type (i,i+3) in our data set consisting of 4391 globular proteins. In all the panels, the blue curves depict the histograms for embedded poking contacts, whereas the green curves show the corresponding histograms for isolated poking contacts. The mean values and standard deviations are indicated in a color-coded manner. The vertical red lines indicate the theoretically predicted value of a given geometrical attribute for the Kepler helix. a) shows the distribution of the (i,i+3) distances. The vertical red line indicates the theoretically predicted coin diameter of 2Δ = 5.26Å, pertaining to Kepler touching. b) shows the distribution of the angles (i-1,i,i+3) and (i,i+3,i+4) that are predicted to be 90° (shown as the vertical red line) for embedded poking contacts. c) shows the distributions of the scalar products $\mathbf{t}_i \cdot \mathbf{b}_{i+3}$ and $\mathbf{t}_{i+3} \cdot \mathbf{b}_i$. The vertical red line indicates the theoretically predicted value of +0.241 (corresponding to an angle ~ 76°) for the Kepler helix. Note that positive values correspond to a right-handed helix, while negative values correspond to a left-handed helix. The embedded (i,i+3) poking contacts are constrained to be right-handed (because of local steric clashes along the backbone comprised of left-handed amino acids). However isolated poking contacts do not have to satisfy this requirement. d) shows the histograms of the scalar products $\mathbf{t}_i \cdot \mathbf{r}_{i,i+3}$ and $\mathbf{t}_{i+3} \cdot \mathbf{r}_{i,i+3}$. The vertical red line indicates the theoretically predicted value of +0.213 (corresponding to an angle ~ 77.7°) for the Kepler helix. Here $\mathbf{r}_{i,i+3}$ is the vector joining i with i+3.

## 3.3 Kepler Sheets

To characterize handholding in Kepler sheets, we have analyzed 13,442 parallel poking contacts and 28,118 antiparallel ones of the type (i,j) with j>i+3. In a Kepler helix, an interior site i has two poking contacts with sites i+3 and i-3. A canonical strand, on the other hand, has no poking contacts within itself and therefore all constraints are imposed by neighboring strands within a sheet. We eliminate edge effects by choosing contacts in the interior of the sheet by requiring that an (i,j) poking contact in a parallel arrangement is accompanied by the pairs of beads (i-2,j-2) and (i+2,j+2) also being poking contacts. For the antiparallel arrangement, the (i,j) poking contact is required to have partner pairs of beads (i-2,j+2) and (i+2,j-2) that are poking contacts as well. For the parallel case (Figure 10a), the coupled axes are at a distance consistent with the coin diameter $2\Delta$, whereas the two axes that are independent of each other tend to squeeze closer to each other. This degree of freedom is afforded to the sheet structure unlike the more rigid Kepler helix. A similar effect is observed for an antiparallel arrangement in Figure 10b. Figure 10c is a depiction of histograms of certain angles (defined in the caption) for parallel and antiparallel sheets that ought to be 90°, according to the theory of Kepler handholding in sheets. The data are in approximate accord with theoretical expectations [1]. The local planarity of the strands is shown in Figure 10d as measured by the dihedral angle µ. The results are in good accord with the expectation that there is a weak left-handed twist (µ is consistently greater than 180° for the strands whereas it is around 60° for the right-handed helix) in the strands, while yet allowing a strand to be locally approximately planar. Figure 10e shows pictorially the consistent zigzagging of strands with virtually the same behavior in both parallel and antiparallel arrangements. Finally, Figure 10f depicts a vivid example of flexibility in the antiparallel sheet with a bimodal histogram of **$n_i$·$r_{ij}$** and **$n_j$·$r_{ij}$** scalar products reflecting the difference between on- and off-axes (i,j) pairs.

Around ~ 96% of the (i,j) non-local poking interior sheet contacts correspond to a DSSP [22] categorization of (E,E), while around ~2% of such pairs are classified as (C,C) pairs by DSSP. About 0.8% of the poking sheet contacts correspond to (E,C) and (C,E) DSSP contacts. Most of the embedded non-local poking contacts are associated with a hydrogen bond. Zig-zag hydrogen bonds characteristic of parallel strands are associated with about ~99.5% of poking contacts, while the ladder hydrogen bonding pattern characteristic of antiparallel strands are associated with more than ~92% of the corresponding poking contacts. This provides additional evidence for the remarkable marriage between geometry and chemistry [30].

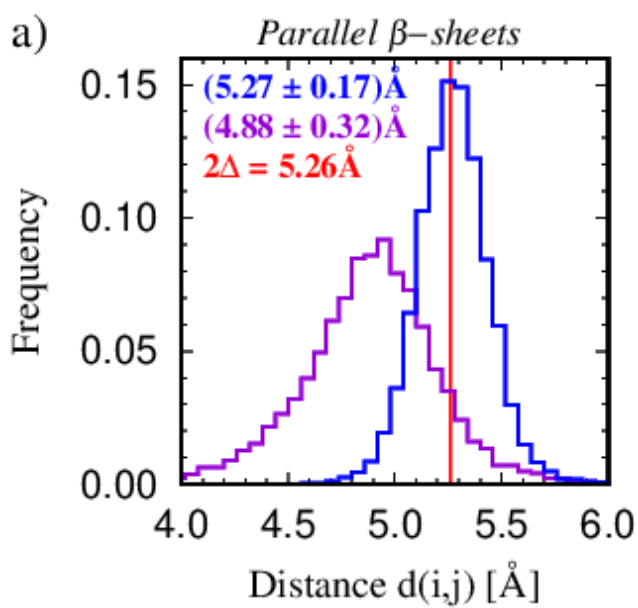
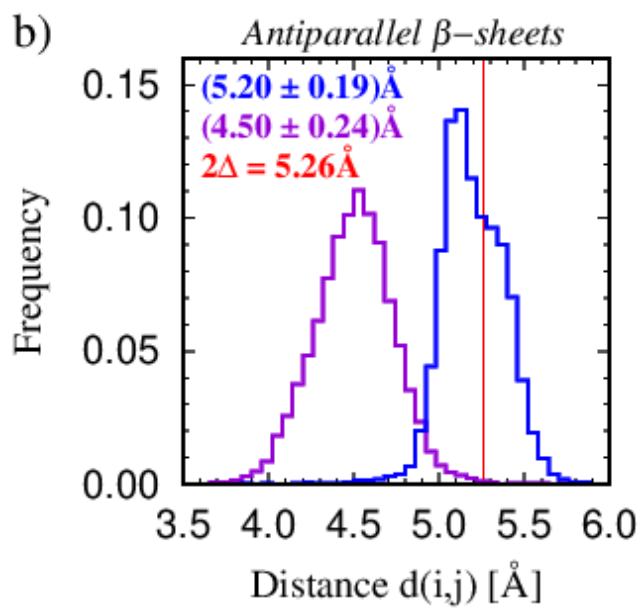
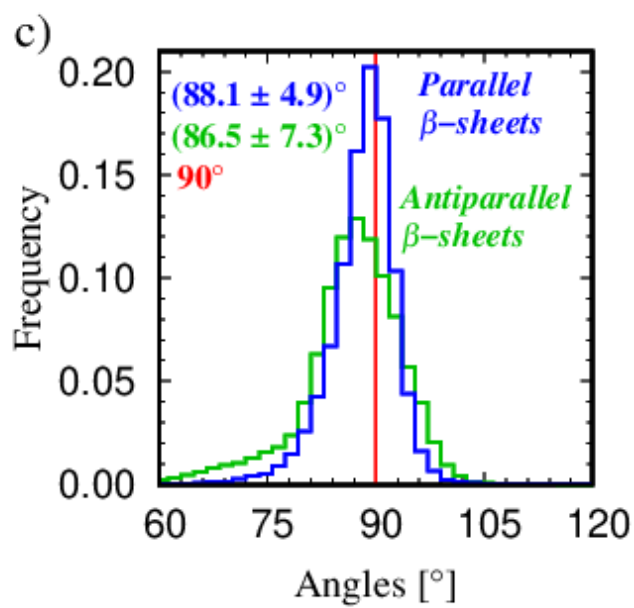
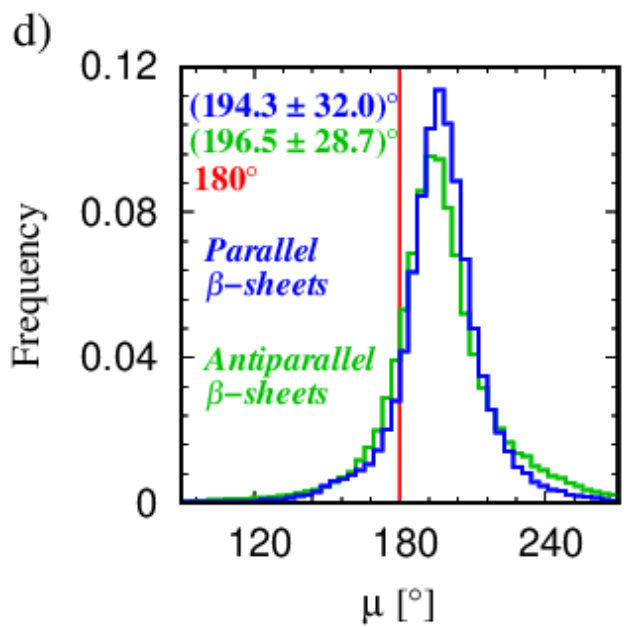
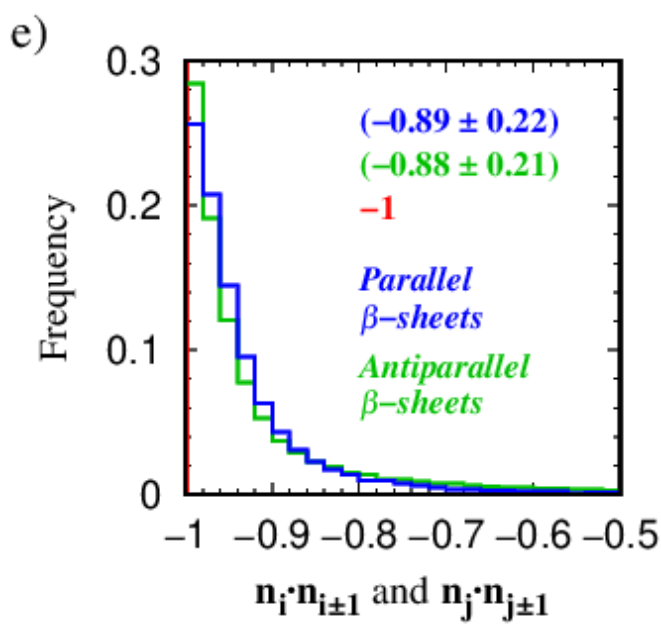
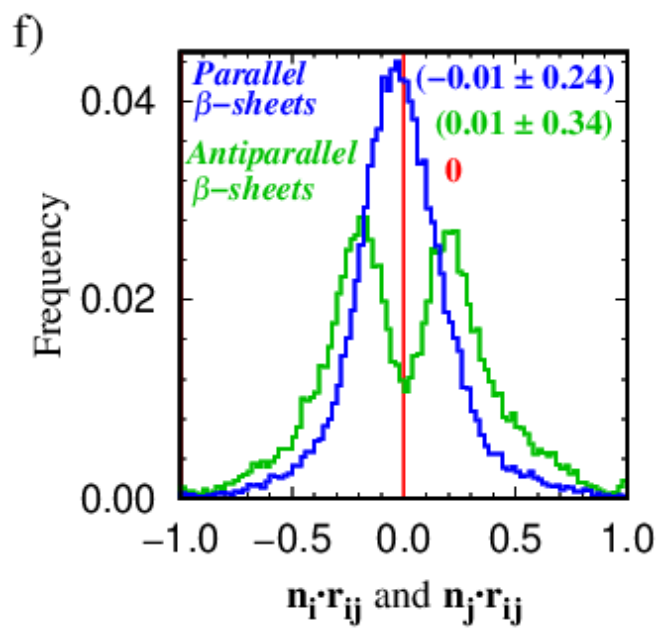

Figure 10: Geometrical characteristics of non-local poking contacts in parallel and antiparallel arrangements within a sheet. a) shows the histograms of the (i,j) distances in the parallel arrangement of chain segments for i and j on coupled axes (in blue) and off the coupled axes (in purple). In the coupled axes case, the two axes are theoretically predicted to be a distance 2Δ apart to allow for Kepler coin touching. In contrast, the axes that are not paired can come closer together and facilitate compaction. The vertical red line indicates the theoretically predicted distance of 2Δ = 5.26Å. The mean values and standard deviations are indicated. b) is a histogram for the antiparallel case showing the distribution of the distances (i,$M_j$) and ($M_i$,j) for a pair of points lying on the coupled axes (in blue) and off the coupled axes (in purple). c) shows the distribution of the angles (i±2,i,j) and (i,j,j±2), predicted to be 90° in the antiparallel arrangement (in blue). For the parallel arrangement (in green), the corresponding angles, (i±1,i,$M_j$) and (i,$M_j$,j±1), need to be 90° when (i,$M_j$) are on the paired axes and the angles (i±1,$M_i$,j) and ($M_i$,j,j±2) ought to be 90° when instead the points ($M_i$,j) are on the paired axes. d) shows the distribution of the dihedral angle of the segment (i,i+1,i+2,i+3) defined as the angle between the planes defined by (i,i+1,i+2) and (i+1,i+2,i+3). The vertical red line indicates the value of 180°, the dihedral angle of the idealized zig-zag strand. e) shows the distributions of the cosines of angles between successive normals along the zig-zag strands. The vertical red line indicates the value of -1 corresponding to the angle of 180° between successive normals in an idealized zig-zag strand. f) shows the distributions of scalar products ($n_i$, $r_{ij}$) and ($n_j$,$r_{ij}$). The vertical red line indicates the idealized value of the scalar product of 0. The antiparallel case exhibits a bimodal histogram, which reflects the squeezing leading to compaction of the off-axis pair of points.

## 3.4 Computer Simulations

The computer simulation model is described in detail in our earlier paper [1] and in the Materials and Methods Section. To recapitulate, we assign *all* poking contacts within 12Å with a reward of -$E_\gamma$. We consider two special cases of poking contacts (i,j) within 6Å for an extra reward. The first case involves (i,i+3) contacts with soft constraints on the values of the dot products, **$t_i \cdot b_{i+3}$**, **$b_i \cdot t_{i+3}$** and **$t_i \cdot r_{i,i+3}$** and **$t_{i+3} \cdot r_{i,i+3}$**, and are allocated an additional -($E_\alpha$ - $E_\gamma$) reward. The second involves (i,j) contacts with soft constraints on the dot products **$n_i \cdot n_{i-1}$**, **$n_i \cdot n_{i+1}$**, **$n_j \cdot n_{j-1}$**, **$n_j \cdot n_{j+1}$**, **$n_i \cdot r_{ij}$**, and **$n_j \cdot r_{ij}$** and are rewarded additionally by an amount -($E_\beta$ - $E_\gamma$). The total energy of a conformation is simply given by $E = -E_\alpha N_\alpha - E_\beta N_\beta - E_\gamma N_\gamma$ (with $E_\alpha = E_\beta = 1$), where $N_\alpha$, $N_\beta$, and $N_\gamma$ denote the number of helix-like, sheet-like, and assembly promoting contacts in each conformation.

In our simulations, we have explored the values of $E_\gamma$ from 0 to 0.5. We observe dynamical switching between different structures in Monte Carlo trajectories at low temperatures in the RE approach. From these simulations, we have selected 36 distinct conformations belonging to three topological families. A gallery of 12 distinct conformations each in the all-α, all-β, and α+β categories are presented in Figure 11. Supplementary information contains a video showing dynamical switching between conformations of different topologies for a chain of length 80 over a 8000 frame long Monte Carlo trajectory at low-temperature ($T^*$=0.05), in which successive frames were recorded at time intervals of 5000 Monte Carlo steps per bead.

In our simulations, we have one free energy parameter $E_\gamma$, which drives compaction. When $E_\gamma = 0$, there is no drive to promote tertiary structure formation and independent helical and sheet conformations are obtained. A non-zero attractive parameter $E_\gamma$ promotes the assembly of secondary motifs into protein-like tertiary structures of different topologies. When $E_\gamma \sim 0.35$, the drive to form secondary motifs is reduced and the building blocks themselves start to become unstable. For each of the 36 conformations, we calculate energies $E = -E_\alpha N_\alpha - E_\beta N_\beta - E_\gamma N_\gamma$ (with $E_\alpha = E_\beta = 1$) as a function of $E_\gamma$. We define a degeneracy parameter Q as the ratio of the standard deviation of the 36 energies to the magnitude of the mean energy value. Figure 12 is a plot of Q versus $E_\gamma$. It exhibits a broad minimum around $E_\gamma \sim 0.22$ underscoring the robustness of the near degeneracy of the energies of distinct conformations, even for a homopolymer chain of modest length.

Figure 13 shows the energies of the 36 conformations shown in Figure 11 and labelled along the x-axis for the energy parameters $E_\alpha=1$, $E_\beta=1$, and $E_\gamma=0.22$. Note that all-α configurations completely lack contacts in the β-basin whereas all-β configurations have no contacts in the α-basin. α+β conformations contain both. The three energy parameters conspire to yield near degeneracy in the energies of all distinct topologies. Also, there needs to be some approximate tuning of $E_\gamma$ to promote assembly of the building blocks of helices and sheets, while yet not destabilizing them. This illustrates how the presculpted energy landscape [20] of proteins is captured by our model in a simple way.

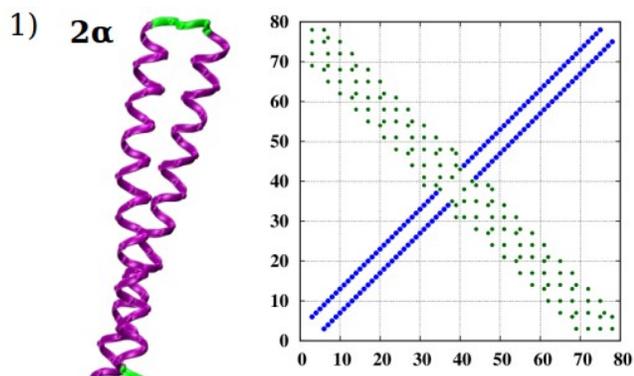
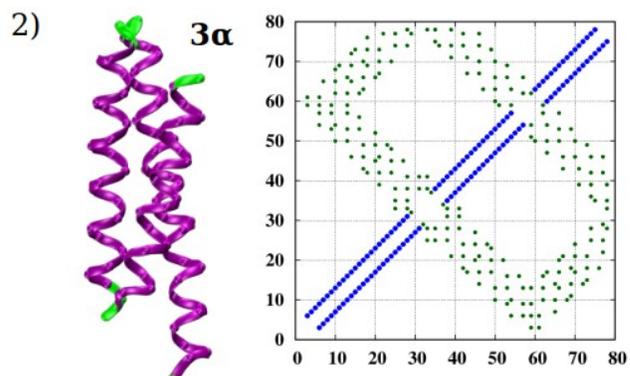
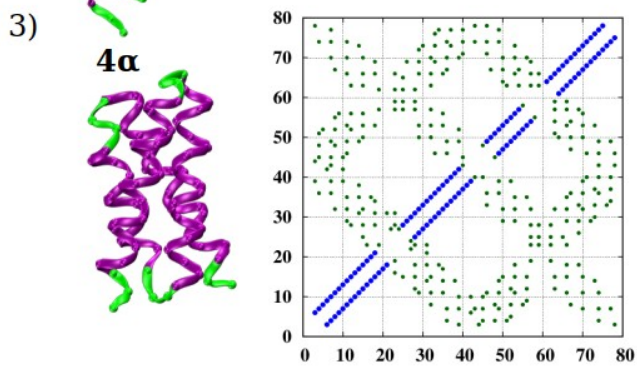
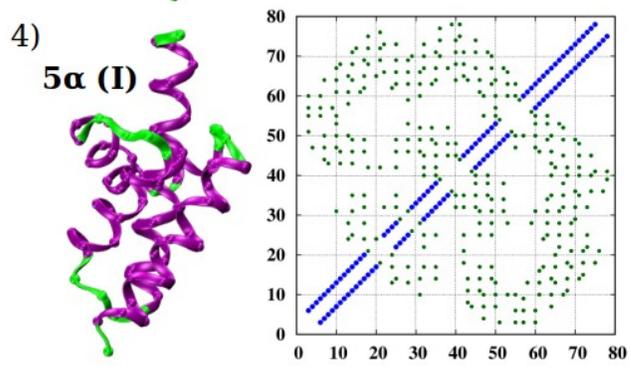
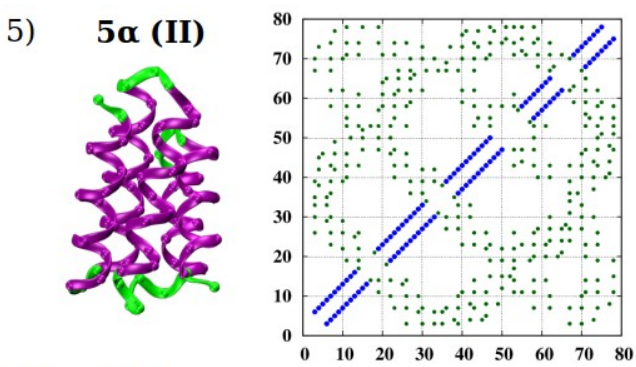
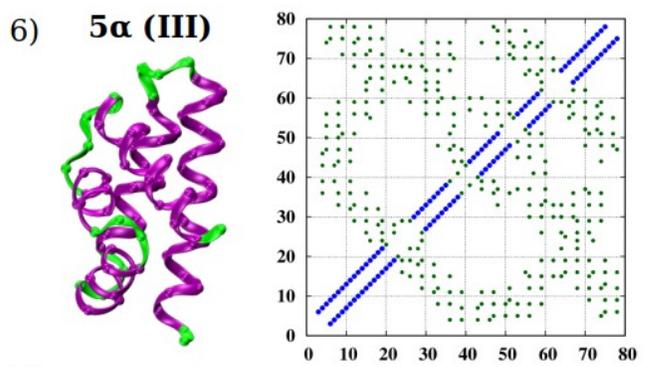
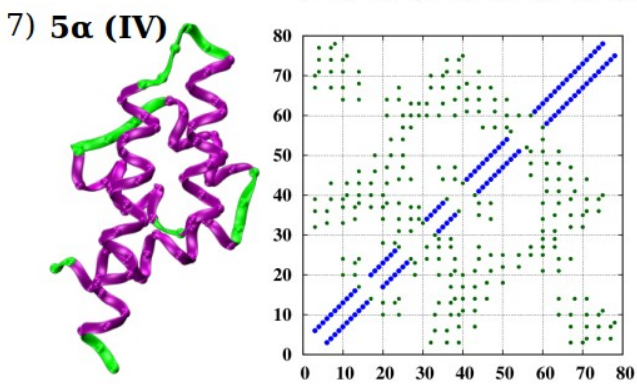
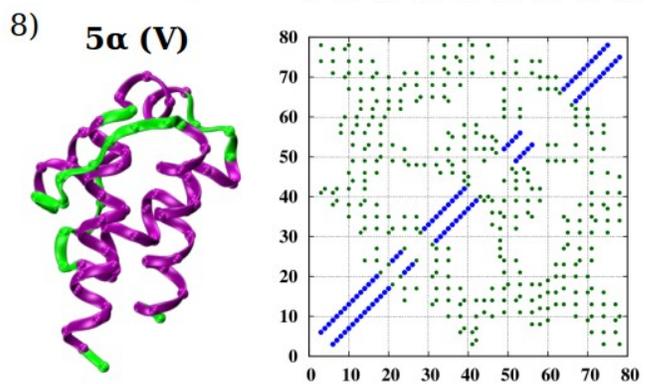

9) **6α (I)** 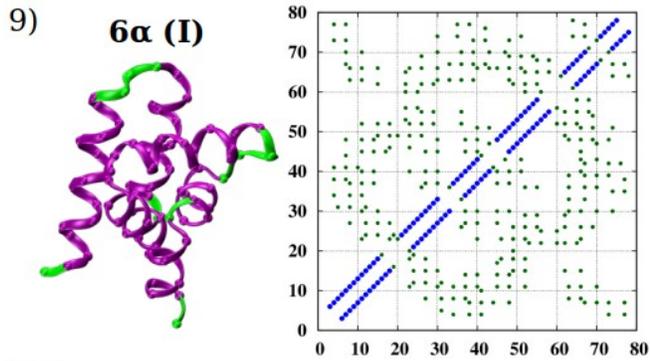

10) **6α (II)** 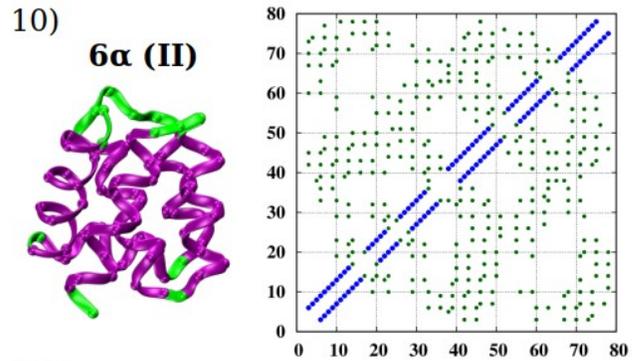

11) **6α (III)** 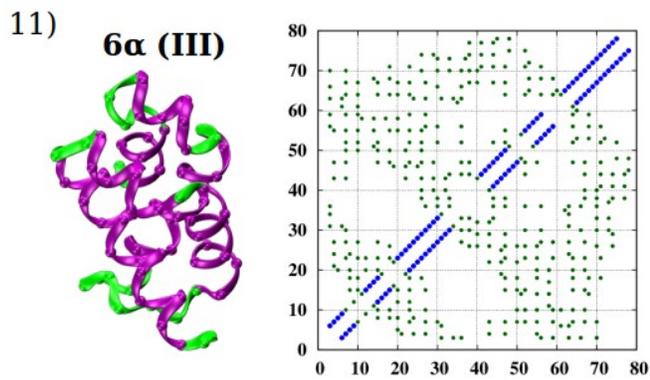

12) **7α** 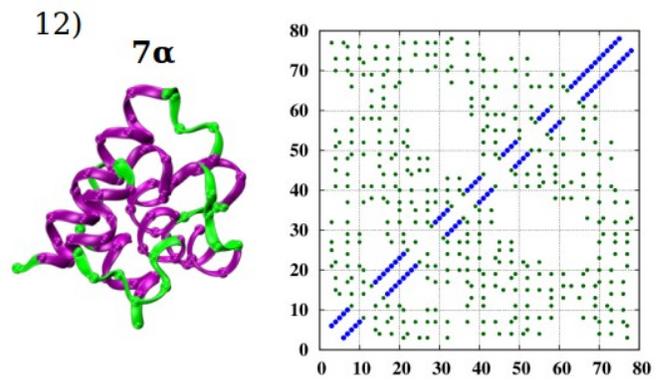

13) **β-barrel (I)** 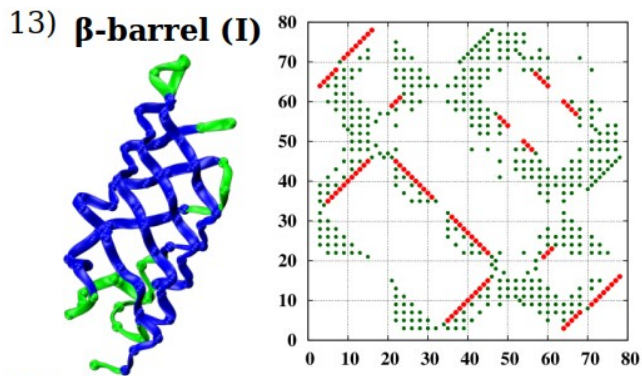

14) **β-barrel (II)** 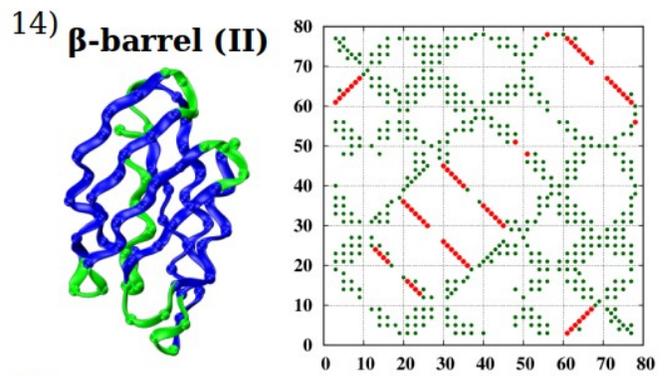

15) **β-barrel (III)** 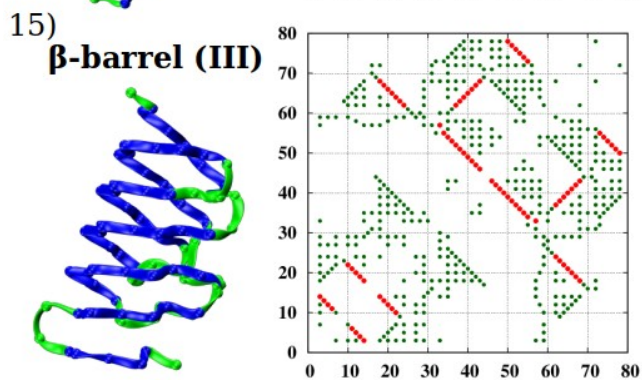

16) **β-barrel (IV)** 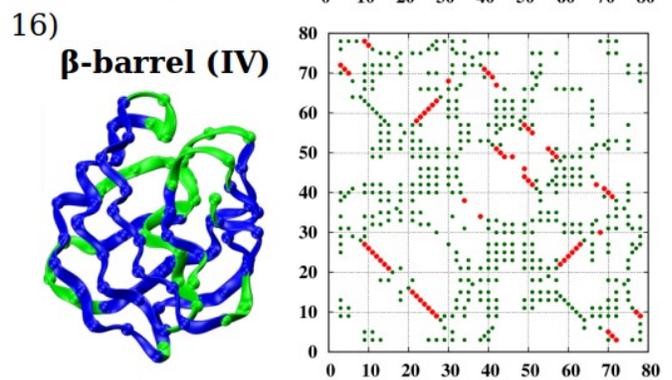

17) **2β-sheets (I)** 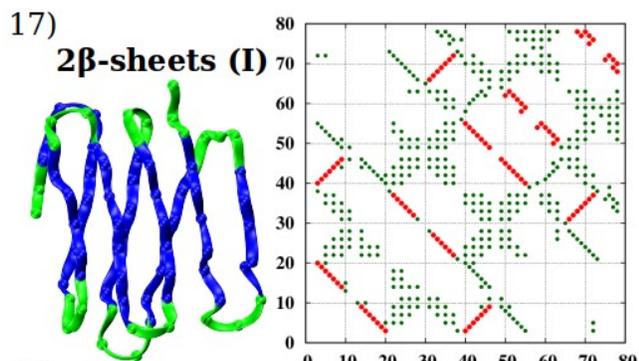

18) **2β-sheets (II)** 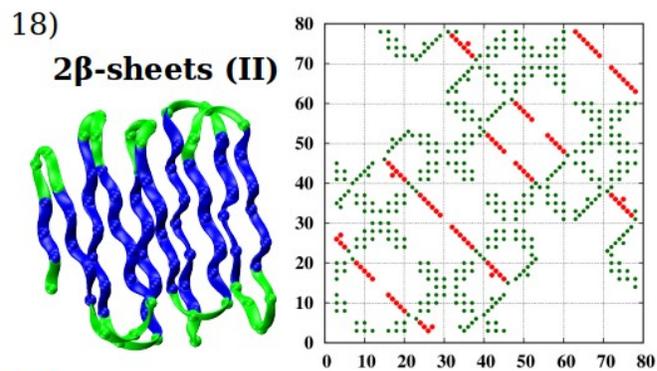

19) **2β-sheets (III)** 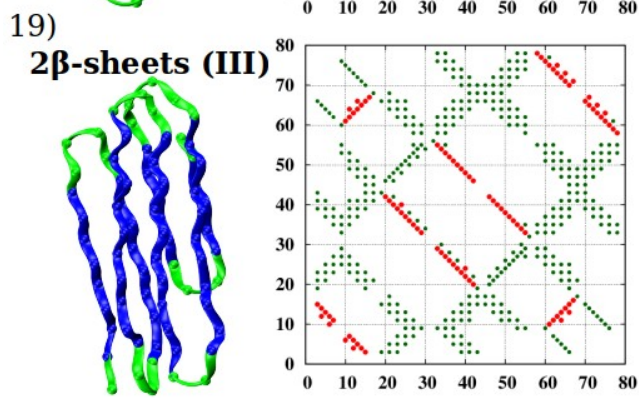

20) **2β-sheets (IV)** 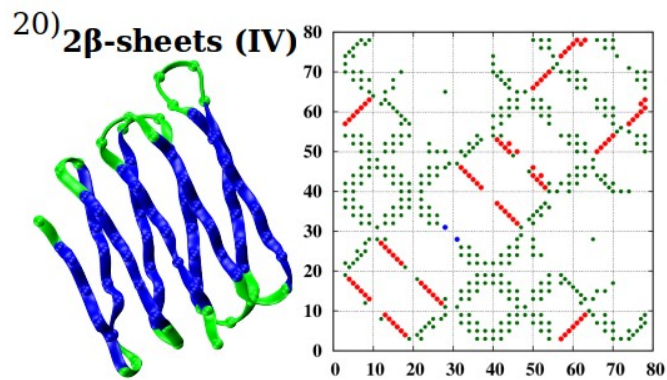

21) **2β-sheets (V)** 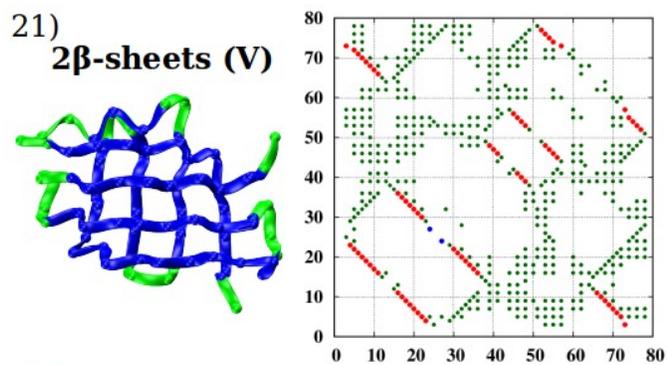

22) **2β-sheets (VI)** 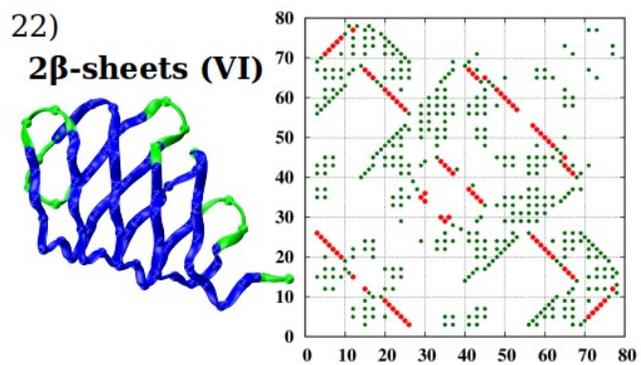

23) **3β-sheets (I)** 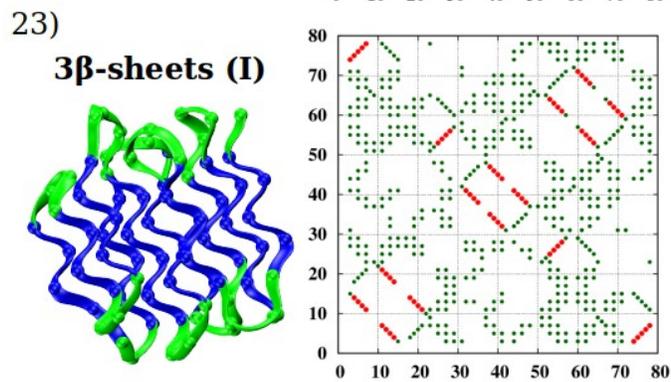

24) **3β-sheets (II)** 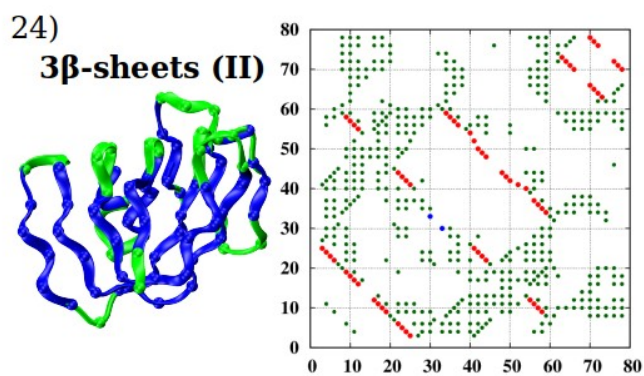

25) **α+4β** 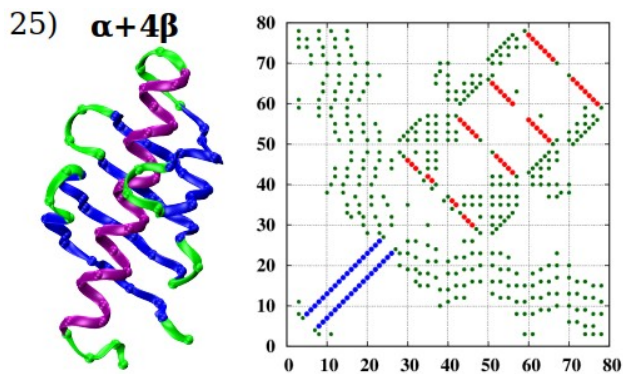

26) **α+5β** 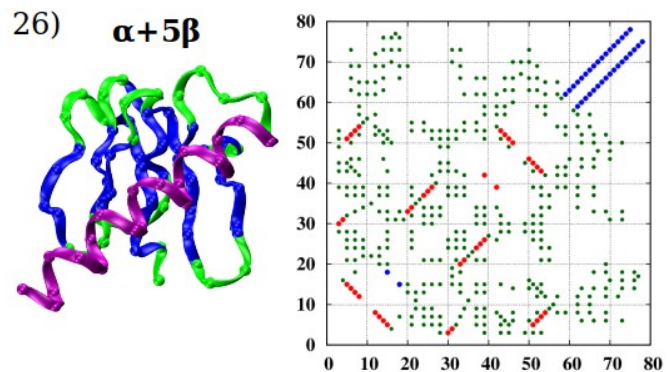

27) **α+6β** 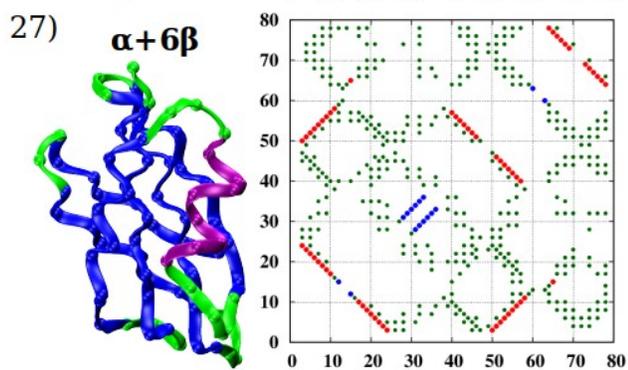

28) **α+8β** 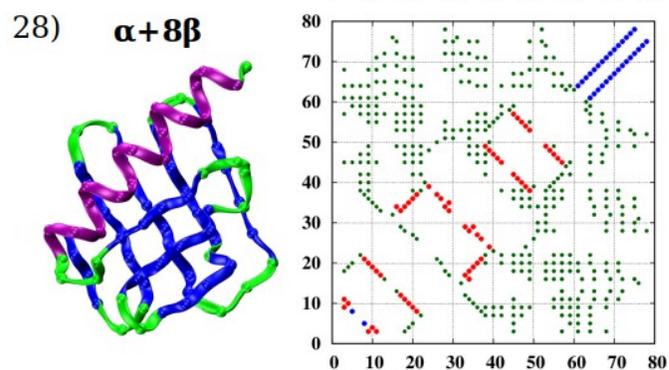

29) **2α+5β** 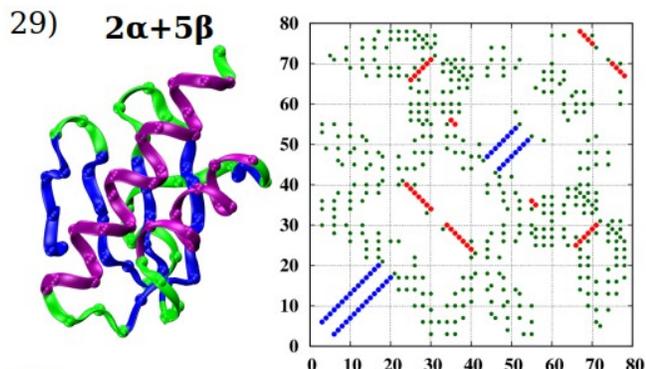

30) **2α+6β** 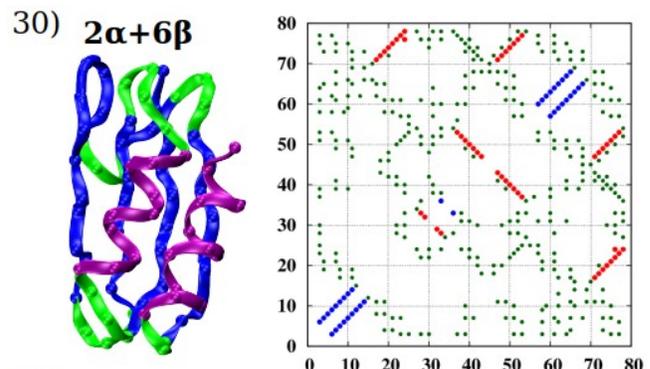

31) **3α+3β (I)** 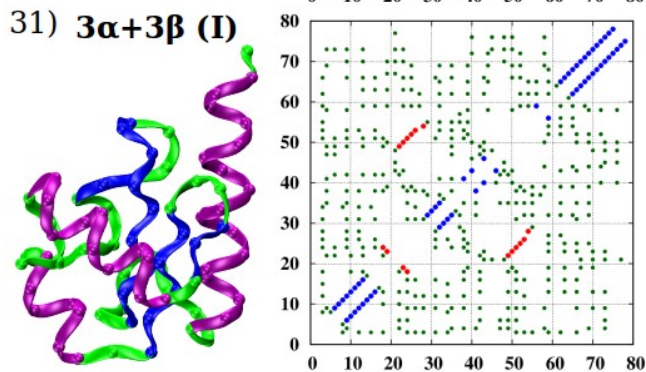

32) **3α+3β (II)** 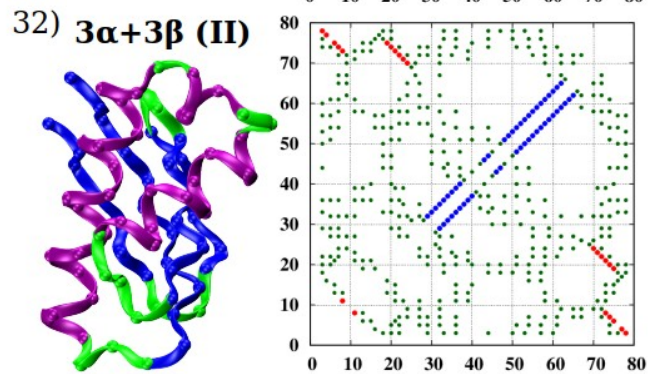

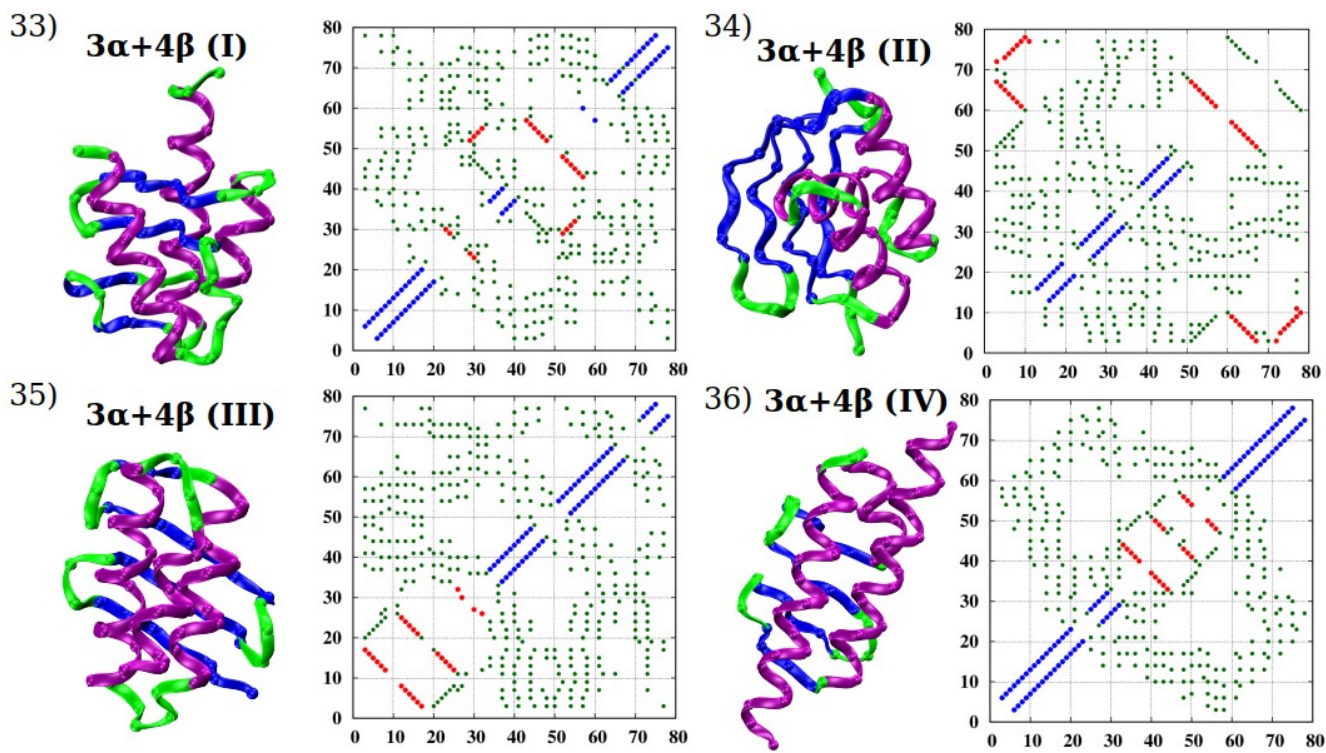

Figure 11: Gallery of 36 low-lying states of different topologies extracted from the numerical simulations of a chain of length 80, along with their contact maps. The structures are drawn in ribbon representation, with helices shown in purple, β-strands in blue, and loops in light green color. The positions of $C_\alpha$ atoms are shown as spheres. In the contact maps, the blue points represent the contacts that belong to the helical basin, the red points represent contacts in the sheet basin, while the dark green points (shrunk in size for clarity) denote the poking contacts within 12Å that promote assembly. The conformations 1-12 have an all-α topology, 13-16 have an all-β topology (forming a β-barrel), 17-24 also have an all-β topology (but this time made of two or three β-sheets), and 25-36 have an α+β topology.

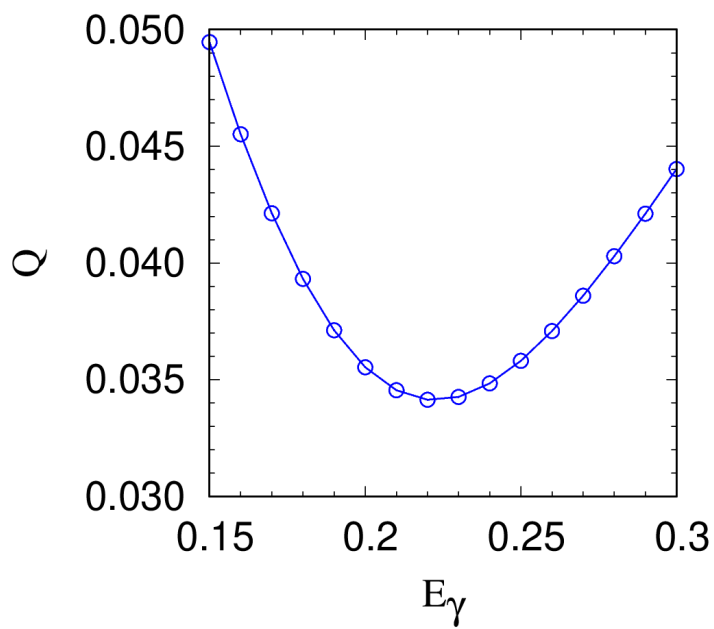

Figure 12: Plot of Q, quantifying the degree of degeneracy of the energies of the 36 conformations shown in Figure 11 (see text), versus the magnitude of the energy parameter promoting tertiary assembly of the secondary motifs, $E_\gamma$. Note that Q is less than 0.05 for the entire range of the energy parameter in the figure.

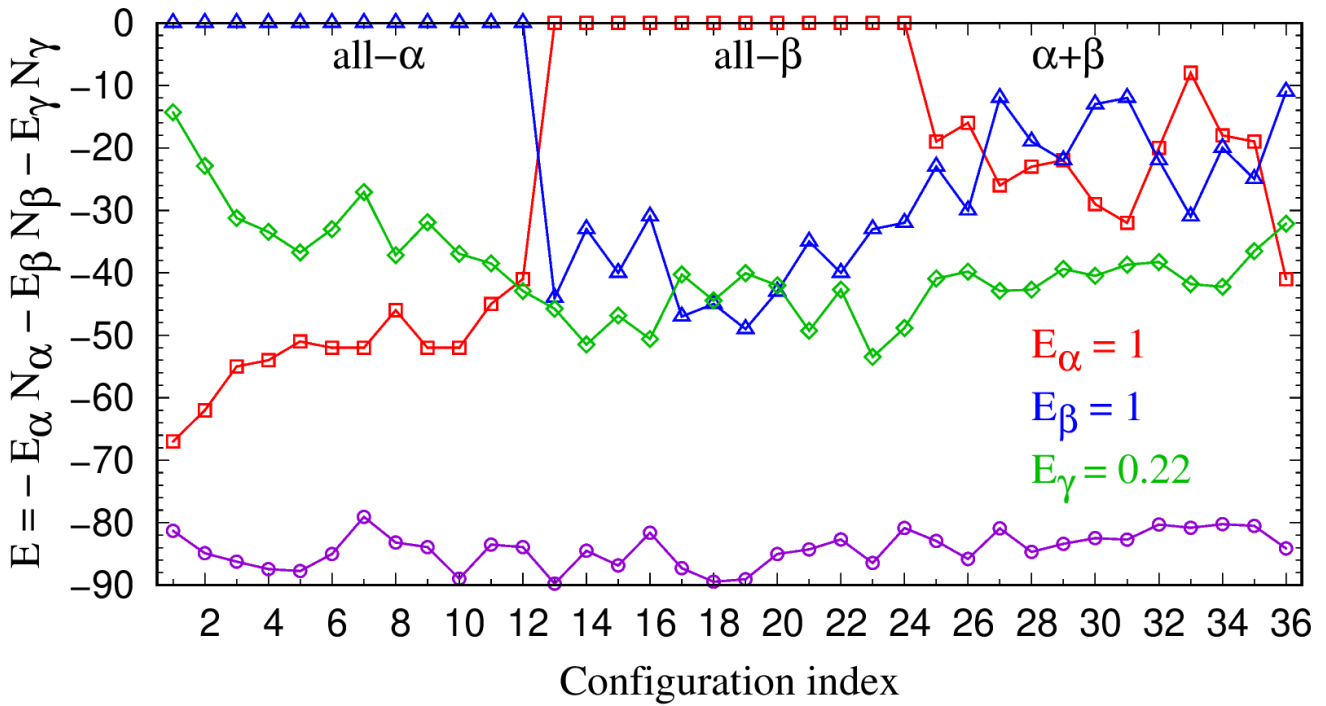

Figure 13: Plot of the energies of the 36 conformations shown in Figure 11, $E=-E_\alpha N_\alpha -E_\beta N_\beta -E_\gamma N_\gamma$ with $E_\alpha=E_\beta=1$ and $E_\gamma=0.22$. Q exhibits a minimum around this value of the assembly energy parameter indicating good degeneracy (Figure 12). The blue triangles denote the 'helical' contribution to the energy, the red squares indicate the 'sheet' contribution, and the green diamonds denote the 'assembly' contribution. The purple circles denote the total energy of each of the 36 conformations underscoring the near degeneracy.

## 4. Conclusions

Globular proteins [14-17] act as machines in living cells, and they serve as the molecular targets of evolution. Proteins, despite their moderate size, exhibit many common characteristics. Small globular proteins fold rapidly and reproducibly into their native state structures. The native state folds do not undergo mutations unlike sequences. All protein native state structures are made of the same building blocks: topologically one-dimensional helices and zig-zag strands assembled into almost planar sheets. The number of protein sequences vastly exceeds the number of distinct native state folds, and many sequences can adopt the same native state conformation. The native state structure of a protein is generally robust to amino acid mutations except at certain pivotal locations. Also, several proteins tend to aggregate creating water-insoluble amyloid [31-33]. These commonalities can be understood based on our geometrical model of proteins.

Symmetry and geometry constrain the structures of infinite sized crystals with exactly 230 distinct space groups in 3 dimensions [34]. Proteins are modest sized chains. Our analysis strongly suggests that space-filling and symmetry result in the number of distinct folds being of the order of several thousand. Just as common salt and a grocer's apples adopt the face-centered-cubic lattice structure, different sequences of proteins can be housed in the same protein fold [35,36]. The key point is that a putative protein native state fold transcends the sequence housed in it and is determined by the overarching constraints of geometry and symmetry. Indeed, many protein sequences adopt the same fold [35,36], and the menu of potential folds is limited in number [37-41]. In this respect, proteins exhibit some similarities to crystals because of space-filling even though proteins are neither infinitely long nor periodic.

Our work reinforces earlier findings [11,20,42-45] that the gross features of the energy landscape of proteins result from the amino acid *aspecific* common features of all proteins. This landscape is *(pre)sculpted* by geometry and symmetry (and not by chemistry) and has several thousand broad minima corresponding to putative native state structures. For each of these minima, the desired funnel-like behavior [46-48] is already achieved at the *homopolymer* level. The superior fit of a sequence to one of the folds from the predetermined menu sculpts the second and simpler stage of the folding funnel landscape.

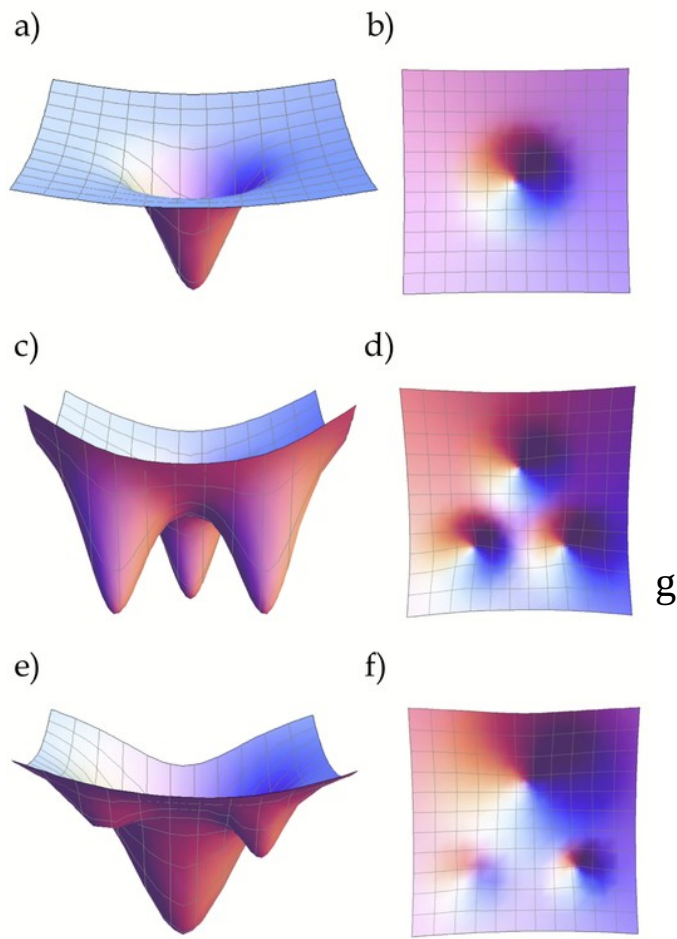

Figure 14: Cartoon of free energy landscape for protein folding. Panels a) and b) show two views of the generally accepted folding funnel landscape of proteins. Panels c) and d) depict two views of an energy landscape with three minima. The number of such minima, predicted by our theory, is several thousand corresponding to the number of topologically distinct ways of assembling the building blocks of helices and sheets. This presculpted landscape is then modified in panels e)

and f) by a sequence designed through evolution as it selects its best fit native state fold.

Protein structures necessarily lie in a marginally compact phase because the tube diameter and the length scale associated with the touching of two coins are self-tuned to be identical. When the tube diameter is larger than the interaction range, the tube cannot avail of any attraction, and one obtains a highly degenerate self-avoiding swollen phase. When the tube thickness is much smaller than the interaction range, one obtains a highly degenerate compact phase with a great deal of latitude in the relative placement of nearby tube segments. Nestled in between is the marginally compact phase, in which there is a great reduction in the degeneracy of the ground state structures with a requirement that nearby tube segments be right alongside tracking each other. This marginally compact phase is poised close to other phases and therefore confers exquisite sensitivity to the structures housed in it.

Proteins are special in that they exhibit stability, diversity, and sensitivity. The individual minima in the free energy landscape are themselves stable and are additionally stabilized by the characteristics of a sequence that fits within it. Diversity exists because there are many low energy equivalent modular structures. Sensitivity stems from the marginally compact phase that protein structures live in.

Our results confirm that there is surprising accord between geometry and quantum chemistry in shaping the structure of proteins. The formidable complexity of proteins seems to be captured by extraordinarily simple ideas from symmetry and geometry. In future work, we will present details and comparisons of our predictions with data on amyloids and investigate the critical role played by side chains.

## Author contributions

TŠ performed the data curation, data analysis, and computer simulations, all with the guidance of JRB. JRB wrote the paper. All authors read and helped improve the manuscript.

## Acknowledgements

We are indebted to George Rose for fruitful collaboration. The computer calculations were performed on the Talapas cluster at the University of Oregon.

## Funding information

This project received funding from the European Union's Horizon 2020 research and innovation program under Marie Skłodowska-Curie Grant Agreement No. 894784 (TŠ). The contents reflect only the authors' view and not the views of the European Commission. JRB was supported by a Knight Chair at the University of Oregon. TXH is supported by The Vietnam Academy of Science and Technology under grant No. NVCC05.05/22-23.

## Conflict of interest

Authors declare that there is no conflict of interest.